\documentclass[11pt]{article}
\newcommand{\Comment}[1]{{}}
\usepackage{amsfonts,amsmath,epsfig,color,latexsym,graphics}
\usepackage{cite}
\Comment{\usepackage{color}
\definecolor{MyDarkBlue}{rgb}{0.15,0.15,0.45}
\usepackage[linktocpage=true]{hyperref}
\hypersetup{
colorlinks=true,
citecolor=MyDarkBlue,
linkcolor=MyDarkBlue,
urlcolor=MyDarkBlue,
pdfauthor={Stephen G. Naculich, Horatiu Nastase and Howard J. Schnitzer},
pdftitle={
},
pdfsubject={hep-th}
}

\usepackage[numbers,sort&compress]{natbib}
\usepackage{hypernat}}

\def\IZ{\relax\ifmmode\mathchoice
{\hbox{\cmss Z\kern-.4em Z}}{\hbox{\cmss Z\kern-.4em Z}}
{\lower.4pt\hbox{\cmsss Z\kern-.4em Z}}
{\lower1.2pt\hbox{\cmsss Z\kern-.4em Z}}\else{\cmss Z\kern-.4em Z}\fi}
\newcommand{\Z}{\mathsf{Z}\kern -5pt \mathsf{Z}}
\newcommand{\unit}{\mathsf{1}\kern -3pt \mathsf{l}}
\def\bone{1\kern -3pt \mathrm{l}}
\def\one{1\kern -3pt \mathrm{l}}
\def\one{{\,\hbox{1\kern-.8mm l}}}

\def\half{\textstyle{1 \over 2}}
\def\third{\textstyle{1 \over 3}}
\def\twothirds{\textstyle{2 \over 3}}
\def\fourthirds{\textstyle{4 \over 3}}
\def\sixth{\textstyle{1 \over 6}}
\def\twentyfourth{\textstyle{1 \over 24}}
\def\hundredtwentieth{\textstyle{1 \over 120}}

\def\cA{  {\cal A}  }

\def\cN{  {\cal N}  }
\def\cO{  {\cal O}  }

\def\bT{\mathbf{T}}

\def\bS{\mathbf{S}}

\def\bGam{\mathbf{\Gamma}}
\def\ones{\left( \begin{array}{c} 1 \\ 1 \\ 1 \\ \end{array} \right) }
\def\Tete{{(2\ell,2 \ell)}}
\def\Tktk{{(2k,2k)}}
\def\Tetemo{{(2\ell,2 \ell-1)}}
\def\Tetemt{{(2\ell,2 \ell-2)}}
\def\Teptep{{(2\ell+1,2 \ell+1)}}
\def\Tkptkp{{(2k+1,2 k+1)}}
\def\Tepte{{(2\ell+1,2 \ell)}}
\def\Teptemo{{(2\ell+1,2 \ell-1)}}
\def\Teptemt{{(2\ell+1,2 \ell-2)}}

\def\lam{\lambda}
\def\Lf{\Lambda_4}
\def\Mf{M_4} 
\def\eps{\epsilon} 
\def\ep{\epsilon}
 
\def\sumz{  \sum_{L=0}^\infty }
\def\sumo{  \sum_{L=1}^\infty }
\def\mommu{ { s_{ij} \over \mu^2 } }
\def\as{a(\mu^2)}
\def\bas{\bar{a}}

\def\pel{{(\ell)}}
\def\pf{{(f)}}
\def\Ell{{(L)}}

\def\EllL{{(L,L)}}
\def\Ellk{{(L,k)}}
\def\Ellodd{{(2\ell+1,2\ell+1)}}
\def\Elleven{{(2\ell,2\ell)}}

\def\Ellzero{{(L,0)}}

\def\Zero{{(0)}}
\def\One{{(1)}}

\def\Oneone{{(1,1)}}
\def\Two{{(2)}}
\def\Three{{(3)}}
\def\Four{{(4)}}
\def\Five{{(5)}}

\def\Twotwo{{(2,2)}}
\def\Threezero{{(3,0)}}
\def\Threeone{{(3,1)}}
\def\Threetwo{{(3,2)}}
\def\Threethree{{(3,3)}}

\def \Tr {\mathop{\rm Tr}\nolimits}
\def\eqn#1{eq.~(\ref{#1})} 
\def\Eqn#1{Equation~(\ref{#1})}
\def\eqns#1#2{eqs.~(\ref{#1}) and~(\ref{#2})}

\def\theequation{\thesection.\arabic{equation}}

\newcommand\ignore[1]{}

\def\ket#1{\left| #1\right\rangle}

\newcommand{\be}{\begin{equation}}
\newcommand{\ba}{\begin{eqnarray}}
\newcommand{\bea}{\begin{eqnarray}}

\newcommand{\ee}{\end{equation}}
\newcommand{\eea}{\end{eqnarray}}
\newcommand{\ea}{\end{eqnarray}}

\newcommand{\nn}{\nonumber}

\begin{document}

\renewcommand{\thefootnote}{\fnsymbol{footnote}}

\makeatletter
\@addtoreset{equation}{section}
\makeatother
\renewcommand{\theequation}{\thesection.\arabic{equation}}

\rightline{}
\rightline{}
   \vspace{0.8truecm}

\begin{flushright}
BRX-TH-663
\\ 
BOW-PH-156
\end{flushright}

\vspace{10pt}


\begin{center}
{\LARGE \bf{\sc 
All-loop infrared-divergent behavior of 
\\[.2cm]
most-subleading-color
gauge-theory amplitudes
}}
\end{center} 
 \vspace{1truecm}
\thispagestyle{empty} \centerline{
{\large \bf {\sc Stephen G. Naculich${}^{a,}$}}\footnote{E-mail address: \Comment{\href{mailto:naculich@bowdoin.edu}}{\tt 
    naculich@bowdoin.edu}},
    {\large \bf {\sc Horatiu Nastase${}^{b,}$}}\footnote{E-mail address: \Comment{\href{mailto:nastase@ift.unesp.br}}{\tt 
    nastase@ift.unesp.br}}, {\bf{\sc and}}
    {\large \bf {\sc Howard J. Schnitzer${}^{c,}$}}\footnote{E-mail address:
                                \Comment{ \href{mailto:schnitzr@brandeis.edu}}{\tt schnitzr@brandeis.edu}}
                                                           }

\vspace{1cm}

\centerline{{\it$^{a}$ Department of Physics}} \centerline{{\it
Bowdoin College, Brunswick, ME 04011, USA}}

\vspace{.5cm}

\centerline{{\it ${}^b$ 
Instituto de F\'{i}sica Te\'{o}rica, UNESP-Universidade Estadual Paulista}} \centerline{{\it 
R. Dr. Bento T. Ferraz 271, Bl. II, Sao Paulo 01140-070, SP, Brazil}}

\vspace{.5cm}
\centerline{{\it ${}^c$ 
Theoretical Physics Group, Martin Fisher School of Physics}} \centerline{{\it Brandeis University, Waltham, 
MA 02454, USA}}

\vspace{1truecm}

\thispagestyle{empty}

\centerline{\sc Abstract}

\vspace{.1truecm}

\begin{center}
\begin{minipage}[c]{380pt}{\noindent 
The infrared singularities of gravitational amplitudes 
are one-loop exact,
in that higher-loop divergences are characterized by the exponential
of the one-loop divergence.
We show that the contributions to SU($N$) gauge-theory amplitudes
that are most-subleading in the $1/N$ expansion are also one-loop exact,
provided that the dipole conjecture holds.
Possible corrections to the dipole conjecture, 
beginning at three loops,
could violate one-loop-exactness, though would still maintain the absence of collinear divergences. 
We also demonstrate a relation between
$L$-loop four-point $\cN=8$ supergravity and 
most-subleading-color $\cN=4$ SYM amplitudes 
that holds for the two leading IR divergences,
$\cO(1/\epsilon^L)$ and $\cO(1/\epsilon^{L-1})$, 
but breaks down at $\cO(1/\epsilon^{L-2})$.

}
\end{minipage}
\end{center}

\vspace{.5cm}

\setcounter{page}{1}
\setcounter{tocdepth}{2}

\newpage

\renewcommand{\thefootnote}{\arabic{footnote}}
\setcounter{footnote}{0}

\linespread{1.1}
\parskip 4pt

\section{Introduction}

The structure of infrared divergences in scattering amplitudes 
of massless particles has been an object of much study over
past decades.
The IR behavior of gravity amplitudes, in particular,
has a remarkable simplicity \cite{Weinberg:1965nx}, 
traceable to the absence of collinear divergences,
and also to the fact that divergences only arise from 
soft graviton exchange between external particles; 
non-abelian-like interactions among virtual gravitons 
ultimately do not contribute to the IR-divergent behavior. 
As a result, in dimensional regularization (in $D= 4 - 2 \eps$ dimensions),
the leading IR divergence at $L$ loops \cite{Dunbar:1995ed,Naculich:2008ew} 
goes as $1 /\eps^L$,
and further, the IR behavior is one-loop exact;
that is, all $L$-loop divergences arise 
from the exponential of the one-loop 
divergence \cite{Weinberg:1965nx,Dunbar:1995ed,Naculich:2008ew,Naculich:2011ry,White:2011yy,Akhoury:2011kq}.

By contrast, the structure of IR divergences of non-abelian
gauge theories is a richer subject;
both collinear and soft divergences appear.
The IR divergences of a gauge theory amplitude can be factored into 
a product of jet functions and a soft function acting 
on an IR-finite hard function \cite{Sterman:2002qn,Aybat:2006mz}.
The soft function depends on a 
soft anomalous dimension matrix $\bGam^\Ell$
at each loop level.
Recently, strong constraints on the form of $\bGam^\Ell$
were derived using soft collinear effective theory \cite{Becher:2009cu}
and Sudakov factorization and momentum rescaling \cite{Gardi:2009qi}.
The simplest solution to these constraints 
is the sum-over-color-dipoles 
formula \cite{Becher:2009cu,Gardi:2009qi,Becher:2009qa},
which essentially states that $\bGam^\Ell$ is proportional to 
$\bGam^\One$ for all $L$.
(This proportionality had previously been established at two loops 
in ref.~\cite{Aybat:2006mz}, 
and conjectured to be true for all $L$ in ref.~\cite{Bern:2008pv}.)
Although departures from the dipole formula are not ruled out 
at three loops and beyond,
the kinematical dependence of such corrections is highly 
constrained 
\cite{Becher:2009cu,Becher:2009qa,Gardi:2009qi,Dixon:2009ur,DelDuca:2011ae,Ahrens:2012qz}.

One can organize the scattering amplitudes of an $SU(N)$ gauge theory
in a combined loop and $1/N$ expansion.
The leading-color (planar) $L$-loop $n$-point amplitude 
$A^\Ellzero$ is proportional to $g^{n-2}  (g^2 N)^L$,
while the subleading-color amplitudes
$A^\Ellk$, with  $k=1, \cdots, L$, 
are down by $1/N^k$ relative to the planar amplitude.
The most-subleading-color amplitudes $A^\EllL$ are independent of $N$. 
While the leading divergence of $L$-loop planar 
gauge theory amplitudes goes as $\cO(1/\eps^{2L})$,
the subleading-color amplitudes $A^\Ellk$
are less divergent, with a leading divergence of $\cO(1/\eps^{2L-k})$
\cite{Naculich:2008ys,Naculich:2009cv,Nastase:2011mx,Naculich:2011my}.
Consequences of the dipole formula for the IR behavior of subleading-color amplitudes
were derived in ref.~\cite{Naculich:2009cv,Naculich:2011my}.

The fact that the most-subleading-color amplitudes $A^\EllL$ only go 
as $\cO(1/\eps^L)$ suggests that they, like gravity amplitudes, 
have no collinear IR divergences, only soft IR divergences. 
In this paper we will explore whether the IR divergences
of most-subleading-color amplitudes also have the property 
of being one-loop exact, as are gravity amplitudes.
We will show that, provided the 
soft anomalous dimension matrices obey the dipole formula, 
this is indeed the case. 
That is, in a given trace basis, 
the IR divergences can be written in terms of the 
exponential of a matrix that describes the one-loop divergence.
We also provide explicit expressions for the IR divergences
of the most-subleading-color four-point amplitude for arbitrary $L$.

Corrections to the dipole formula, if present, begin at three loops.
We compute the $1/N$ expansion of a possible three-loop correction term
to the dipole formula and show that it would affect the 
most-subleading-color three-loop four-point amplitude
$A^\Threethree$ at $\cO(1/\eps)$,
spoiling the one-loop exactness of its IR behavior,
although collinear IR divergences remain absent.

Finally, the similarity between gravity and most-subleading-color 
gauge-theory amplitudes can be used to deduce a relation between
$L$-loop four-point $\cN=8$ supergravity and 
most-subleading-color $\cN=4$ SYM amplitudes 
that holds for the two leading IR divergences,
$\cO(1/\epsilon^L)$ and $\cO(1/\epsilon^{L-1})$, 
but breaks down at $\cO(1/\epsilon^{L-2})$.

In section~\ref{sect-graviton}, we review the
one-loop exactness of gravity amplitudes.
In section.~\ref{sect-IR}, we demonstrate that
most-subleading-color SU($N$) gauge theory amplitudes are similarly
one-loop exact, provided the dipole conjecture holds.
In section \ref{sect-fourpoint}, we derive an expression
for the full IR divergences of the $L$-loop
four-point most-subleading-color amplitude 
in terms of lower-loop amplitudes. 
In section \ref{sect-correction}, we examine the
effect of a possible three-loop correction to the dipole conjecture.
In section \ref{sect-symsugra} we 
deduce a relation between four-point $\cN=8$ supergravity and 
most-subleading-color $\cN=4$ SYM amplitudes.
Various technical details may be found in three appendices.

\section{Infrared divergences of gravity amplitudes  }
\label{sect-graviton}

The pioneering study of the  IR singularities of gravitational theories
by Weinberg \cite{Weinberg:1965nx} showed 
that these are one-loop exact in the sense that all
IR divergences are characterized by the exponential of the one-loop divergence.
Dunbar and Norridge later revisited this issue in the context 
of string theory \cite{Dunbar:1995ed}.
Recently two of us \cite{Naculich:2011ry} 
reformulated this problem in analogy with the modern treatment of 
IR singularities in gauge theories,
and several other authors have studied and
extended the subject from this point of view 
\cite{White:2011yy,Akhoury:2011kq,Miller:2012an,Beneke:2012xa}.

In ref.~\cite{Naculich:2011ry} it was proposed that the $n$-graviton
scattering amplitude can be written as
\be
A_n  =  S_n  \cdot H_n
\label{factor}
\ee
where $S_n$ is the gravitational soft function,
an IR-divergent factor describing the effects of 
soft graviton exchange between the $n$ external particles,
and $H_n$ is the IR-finite hard function.
Contrary to gauge theories,
there are no jet functions, as collinear singularities are absent 
after summing over diagrams.  
We expand each of the quantities in \eqn{factor} in a loop expansion
in powers of 
$\lambda = (\kappa/2)^2 (4\pi e^{-\gamma} )^\eps$,
where $\kappa^2 =32 \pi G$:
\be
A_n = \sumz   A_n^\Ell\,,
\qquad\qquad
S_n = 1 + \sumo   S_n^\Ell\,,
\qquad\qquad
H_n = \sumz   H_n^\Ell \,.
\ee
IR divergences are regulated using dimensional regularization
in $D= 4 - 2 \eps$, with $\eps < 0$.
Then, due to the fact that all IR singularities are associated
with single graviton exchanges between pairs of external particles,
the soft function is given by the exponential of the one-loop IR divergence
\cite{Weinberg:1965nx,Dunbar:1995ed,Naculich:2008ew,Naculich:2011ry,
White:2011yy,Akhoury:2011kq}
\be
S_n = \exp \biggl[  {\sigma_n\over\epsilon} \biggr],
\qquad \sigma_n =   {\lambda \over 16 \pi^2} 
\sum_{j=1}^n \sum_{i<j}  
s_{ij} \log \left( -s_{ij} \over \mu^2 \right), 
\qquad s_{ij} = (k_i + k_j)^2 .
\label{soft}
\ee
(Any IR-finite contributions from these exchanges 
can be absorbed into $H_n$.)
Hence the IR divergences of the gravitational amplitude
are one-loop exact
\be
A_n  =  
\exp \biggl[  {\sigma_n\over\epsilon} \biggr]
H_n
\label{oneloopexact}
\ee
and the $L$-loop amplitude can be expressed as 
\be
A_n^\Ell = \sum_{\ell = 0}^{L} {1 \over (L-\ell)!}  
\biggl[ {\sigma_n \over \epsilon} \biggr]^{L-\ell} H_n^\pel (\eps),
\label{Lgrav}
\ee
that is, the $L$-loop IR divergences are determined by $\sigma_n$ together
with the IR-finite contributions 
(including terms proportional to positive powers of $\eps$)
of all the lower-loop amplitudes.
By keeping the first two terms
\be 
A_n^\Ell =
{1 \over L!}  \biggl[ {\sigma_n \over \epsilon} \biggr]^{L} H_n^\Zero(\eps)  
+ {1 \over (L-1)!}  \biggl[ {\sigma_n \over \epsilon} \biggr]^{L-1} H_n^\One(\eps)  
+ \cO(1/\eps^{L-2} )
\ee
we observe that the two leading IR divergences of the $L$-loop amplitude
are completely determined by the tree and one-loop amplitudes.
Moreover, since $H_n^\One( L\eps) = H_n^\One (\eps) + \cO(\eps)$,
we see that the two leading divergences of the $L$-loop amplitude 
can be related to the one-loop amplitude evaluated in $D=4-2L\eps$
dimensions:
\be 
A_n^\Ell (\eps)=
 {1 \over (L-1)!}  \biggl[ {\sigma_n \over \epsilon} \biggr]^{L-1} 
A_n^\One(L\eps)  
+ \cO(1/\eps^{L-2} )\,.
\label{Lloopproponeloop}
\ee
In sec.~\ref{sect-fourpoint}, we will find an analogous relationship
for the most-subleading-color YM amplitude.

For the remainder of this section, 
we restrict ourselves to the four-point amplitude of $\cN=8$ supergravity. 
In this case, the all-loop-orders amplitude
is proportional to the tree-level amplitude \cite{Bern:1998ug,Bern:1998sv},
allowing us to define the helicity-independent ratios
\be
M_4  = {A_4 /  A_4^\Zero},
\qquad\qquad
M_4^\pf  = {H_4 /  A_4^\Zero} \,.
\ee
Then \eqns{soft}{oneloopexact} imply
\be
M_4 = \exp \biggl[  { \sigma_4 \over \epsilon } \biggr] M_4^\pf,
\qquad\qquad
\sigma_4 = {\lambda \over 8 \pi^2} \left[ 
s \log \left( -s \over \mu^2 \right)
+t \log \left( -t \over \mu^2 \right)
+u \log \left( -u \over \mu^2 \right)
\right]
\ee
where $s=s_{12}$, $t=s_{14}$, and $u=s_{13}$.
Consequently,
the logarithm of the ratio of the full amplitude to the tree amplitude
\be
\Lf \equiv \log M_4   = {\sigma_4\over \epsilon} + \log \Mf^{(f)}
\ee
only has an IR divergence at one loop
\be
\Lf^\One = \Mf^\One = {\sigma_4\over \epsilon} + \Mf^{(1f)} \,.
\ee
By expanding 
\be
M_4 = 1 + \sumo   M_4^\Ell
 = \exp\left( \Lf \right) = \exp \left( \sumo   \Lf^\Ell  \right) 
\ee
we can obtain explicit expressions at each loop order 
\ba
\Mf^\Two &=& \half (\Lf^\One)^2 + \Lf^\Two \,,
\nn\\ 
\Mf^\Three &=&  \sixth (\Lf^\One)^3 + \Lf^\One \Lf^\Two + \Lf^\Three \,,
\nn\\
\Mf^\Four &=&\twentyfourth  (\Lf^\One)^4 
       + {1\over 2} (\Lf^\One)^2 \Lf^\Two
       + \Lf^\One \Lf^\Three 
        + \half (\Lf^\Two)^2 + \Lf^\Four  \,,
\nn\\
\Mf^\Five &=& \hundredtwentieth (\Lf^\One)^5 
+ \sixth (\Lf^\One)^3 \Lf^\Two 
+ \half (\Lf^\One)^2 \Lf^\Three 
+  \half \Lf^\One(\Lf^\Two)^2 
+  \Lf^\One \Lf^\Four 
+ \Lf^\Two \Lf^\Three 
+ \Lf^\Five \,,
\nn \\ 
&\vdots&
\ea
Since $\Lf^\One$ diverges as $1/\eps$, 
and $\Lf^\Ell$ are  IR-finite for all $L \ge 2$,
we see that the two leading IR-divergent terms of the $L$-loop
amplitude can be expressed in terms of the one-loop amplitude
\be
\Mf^\Ell (\epsilon) =
 {1 \over L!} \left[  \Mf^\One (\epsilon) \right]^L + \cO(1/\eps^{L-2})  \,.
\ee
As in the general case (\ref{Lloopproponeloop}), 
we can also write this as a relation between the $L$-loop amplitude 
and the one-loop amplitude evaluated in $D=4-2L\eps$ dimensions:
\be
\Mf^\Ell (\epsilon) =
{1 \over (L-1)!} \biggl[  { \sigma_4 \over \epsilon } \biggr]^{L-1}
\Mf^\One (L \epsilon) + \cO(1/\eps^{L-2})  \,.
\label{Llooponeloop}
\ee

\section{Infrared divergences of the most-subleading-color YM amplitudes}
\label{sect-IR}

In this section we will explore the IR divergences of $n$-gluon 
amplitudes that are most-subleading in the $1/N$ expansion.
These amplitudes are similar to the $n$-graviton amplitudes discussed
in the previous section in two respects:   
(1) although the leading IR divergence of an $n$-gluon amplitude
at $L$ loops goes as $1/\eps^{2L}$, 
the leading divergence of the most-subleading-color
amplitude is milder, only going as $\cO(1/\eps^L)$, 
due to the absence of collinear divergences, and 
(2) if the dipole conjecture, described below, holds, 
then the IR divergences of the most-subleading-color amplitudes
are one-loop exact; that is,
all IR divergences at $L$ loops are determined by the
exponential of the one-loop IR divergence,
as we will show below.
If the dipole conjecture is not valid, then the first property
(lack of collinear divergences) continues to hold 
but the second does not:  
additional IR divergences unrelated to the one-loop divergence
could be present,
potentially beginning at three loops.     
We will describe the form of a potential three-loop correction
to the most-subleading-color four-point function in sect.~\ref{sect-correction}.

The $n$-point amplitude of particles transforming in the adjoint
representation (e.g., gluons) can be expanded in a trace basis  
$\{ T_\lam \}$,
consisting of single and multiple traces of generators
in the fundamental representation,
\be
\cA = \sum_\lam T_\lam A_\lam
\label{basis}
\ee
where the coefficients $A_\lam$ are referred to as 
color-ordered amplitudes.
It is convenient to organize \cite{Catani:1996jh,Catani:1998bh} 
the color-ordered amplitudes into a 
vector $\ket{A}$.
In an SU($N$) gauge theory,
this vector can be decomposed in a simultaneous loop 
and $1/N$ expansion\footnote{
We have omitted an overall factor of $g^{n-2}$ for an $n$-point function.}
\be
\ket{A} = \sum_{L=0}^\infty \sum_{k=0}^L {\as^L \over N^k } \ket{A^\Ellk}
\label{expansion}
\ee
where 
\be
a(\mu^2) = {g^2(\mu^2) N \over 8 \pi^2} (4 \pi e^{-\gamma} )^\eps 
\ee 
is the 't Hooft coupling
and $\mu$ is the renormalization scale.
Our interest in this paper is in the IR behavior of 
the most-subleading-color amplitudes,
that part of the amplitude that depends only on $g^2(\mu^2)$
with no powers of $N$.  
Hence, we are interested in the terms 
$\ket{A^\EllL}$ in the expansion (\ref{expansion}),
which carry as many powers of $1/N$ as of $\as$.

We follow refs.~\cite{Sterman:2002qn,Aybat:2006mz}
by organizing the IR divergences of a gauge theory amplitude as
\be
\left|   A \left(\mommu,  \as, \ep\right) \right> 
= 
 J \left(\as, \eps \right) \, 
{\bS} \left( \mommu,  \as, \eps\right) 
\left | H \left( \mommu, \as, \eps \right) \right>
\label{YMfactor}
\ee
The prefactors $J$ (``jet function") 
and ${\bS}$ (``soft function") characterize the 
long-distance IR-divergent behavior, 
while the short-distance behavior of the amplitude
is characterized by $\ket{H}$ (``hard function''),
and is finite as $\eps \to 0$.
(Quantities in boldface act as matrices on the color
space vectors.)

The jet function has leading IR behavior 
of $\cO(1/\eps^{2L})$ at $L$-loops
(although the poles of $\log J$ only 
go up through $1/\eps^{L+1}$ in a generic gauge theory
\cite{Aybat:2006mz}, and $1/\eps^2$ in $\cN=4$ SYM theory \cite{Bern:2005iz}). 
The jet function, however, 
is irrelevant to the IR divergences of the most-subleading-color
amplitude because it carries no factors of $1/N$
to accompany the factors of $\as$.

The soft function 
\cite{Sterman:2002qn,Aybat:2006mz}
\be
{\bS} \left( \mommu, \as, \eps \right) 
\,=\,
{\rm P}~{\rm exp}\left[
\, -\; 
\frac{1}{2}\int_{0}^{\mu^2} 
\frac{d\tilde{\mu}^2}{\tilde{\mu}^2}
\bGam \left( \mommu,
         \bas \left(\frac{\mu^2}{\tilde{\mu}^2}, \as, \eps  \right)
       \right) 
\right]\,
\label{YMsoft}
\ee
depends on the 
soft anomalous dimension matrix, which can be expanded as 
\be
\bGam \left( \mommu, \as \right)
= \sum_{L=1}^\infty 
\as^L \ \bGam^\Ell
\left(\frac{s_{ij}}{\mu^2}\right). 
\ee
The one-loop 
soft anomalous dimension matrix is given by \cite{Aybat:2006mz}
\be
\bGam^\One = \frac{1}{N}  
\sum_{j=1}^n \sum_{i<j}  
\bT_i \cdot \bT_j \log \left( {\mu^2 \over -s_{ij}  } \right) 
\label{oneloopanom}
\ee 
where $\bT_i$ are the SU$(N)$ generators in the adjoint representation.
Diagrammatically, the operators $\bT_i \cdot \bT_j$ act 
by attaching a gluon rung between the legs of the $i$th and $j$th external particles.
In terms of the color-ordered expansion (\ref{basis}),
$\bGam^\Ell$ acts on a given element $T_\lam$ of the trace basis
(\ref{basis}) to yield a linear combination
\be
\bGam^\Ell  T_\lambda =
\sum_\kappa T_\kappa  \bGam^\Ell_{\kappa \lambda}
\label{action}
\ee
and it is the matrix $\bGam^\Ell_{\kappa \lambda}$
that then acts on the ket $\ket{H}$.

At this point, we invoke the dipole 
conjecture \cite{Becher:2009cu,Becher:2009qa,Gardi:2009qi},
according to which the 
soft anomalous dimension matrix
$\bGam^\Ell$ is proportional to $\bGam^\One$  for all $L$ 
(with the proportionality constants given by 
the coefficients of the cusp anomalous dimension).
This had previously been proven for $\bGam^\Two$ in ref.~\cite{Aybat:2006mz}, 
and hypothesized to be valid for all $L$ in ref.~\cite{Bern:2008pv}.
Corrections at three loops and above, however, have not (yet) been
ruled out completely, although they are highly 
constrained \cite{Becher:2009cu,Becher:2009qa,Gardi:2009qi,Dixon:2009ur,DelDuca:2011ae,Ahrens:2012qz}.
We assume the validity of the dipole formula for
the remainder of this section, but in sec.~\ref{sect-correction} 
we will consider the possibility of a violation at three loops.

If the dipole conjecture holds, then 
$\bGam^\Ell$ all commute with one another, so 
that path ordering of the exponential 
in \eqn{YMsoft} is irrelevant. 
We can then integrate the terms to obtain 
\be
\bS \left( \mommu, \as, \epsilon \right) 
 = \exp \left[ \sum_{L=1}^\infty 
{\as^L  \over  2 L \epsilon }
{\bGam^\Ell} 
\left(1+ {\cO} \left(\frac{\as}{\epsilon}\right)\right)
\right]
\label{integrated}
\ee
where the leading form of the running coupling is given 
by \cite{Sterman:2002qn,Aybat:2006mz}
\be
{\bas}  \left(   \frac{\mu^2}{\tilde{\mu}^2}, \as, \eps \right) 
=  \as \left( \frac{\mu^2}{\tilde{\mu}^2}   \right)^\eps  
\sum_{n=0}^\infty\left[\frac{\beta_0}{4\pi \epsilon}\left(
\left(     \frac{\mu^2}{\tilde{\mu}^2}   \right)^\eps-1\right)\as\right]^n \,.
\ee
The omitted terms in \eqn{integrated}, which depend on $\beta_0$,
the one-loop coefficient of the beta function,
will not contribute to the most-subleading-color amplitudes
because there are no factors of $1/N$ to accompany the
powers of $\as$.  

Generically, one would expect the 
soft anomalous dimension matrices 
$\bGam_{\kappa\lambda}^\Ell$ to 
contain terms of $\cO(1)$ through $\cO(1/N^L)$ 
in the $1/N$-expansion.
If the dipole conjecture is valid, however,
then $\bGam_{\kappa\lambda}^\Ell$
is proportional to $\bGam_{\kappa\lambda}^\One$,
and hence only contains terms of $\cO(1)$ and $\cO(1/N)$.
Since $\bGam^\Ell$ is multiplied by $\as^L$
but carries at most one power of $1/N$,
only $\bGam^\One$ can contribute to the most-subleading-color 
amplitude,
which now simplifies to 
\be
\ket{A} \bigg|_{\rm most-subleading-color}
 = \exp \left[ \frac{\as}{2\epsilon} 
 \bGam^\One_{\rm sub}   \right]
\ket{H(\eps)}\bigg|_{\rm most-subleading-color}
\label{YMoneloopexact}
\ee
where 
$ \bGam^\One_{\rm sub}$
denotes the $1/N$ contribution of the one-loop 
soft anomalous dimension matrix.

\Eqn{YMoneloopexact} is parallel to the gravitational analog (\ref{oneloopexact}).
It demonstrates that, 
provided the dipole conjecture is valid, 
the IR divergences of the 
most-subleading-color amplitudes are one-loop exact,
that is, determined by the one-loop 
soft anomalous dimension matrix $\bGam^\One_{\rm sub}$
and the finite contributions 
(including terms proportional to positive powers of $\eps$)
of lower loop amplitudes,
just as in the case of gravitational amplitudes.

\section{IR behavior of the most-subleading-color four-point amplitude}
\label{sect-fourpoint}

In the previous section, we showed that, subject to the validity
of the dipole conjecture, the IR divergences of the most-subleading
color amplitudes are one-loop exact, given by 
the exponential of the one-loop 
soft anomalous dimension matrix.  
In this section, we will write the IR divergences of the
most-subleading-color $L$-loop amplitude
explicitly in the case of the four-point function,
using the group-theory relations among four-point
color-ordered amplitudes \cite{Naculich:2011ep}.

For the four-point amplitude, the one-loop 
soft anomalous dimension matrix (\ref{oneloopanom}) becomes
\be
\bGam^\One = \frac{1}{N}  \left[
 \left( \bT_1 \cdot \bT_2 + \bT_3 \cdot \bT_4 \right) \log \left( \mu^2 \over -s\right)
+\left( \bT_1 \cdot \bT_3 + \bT_2 \cdot \bT_4 \right) \log \left( \mu^2 \over -u\right)
+\left( \bT_1 \cdot \bT_4 + \bT_2 \cdot \bT_3 \right) \log \left( \mu^2 \over -t\right)
\right].
\label{GamOneFour}
\ee 
As described in the previous section, to evaluate this operator,
we choose a specific four-point trace basis,
consisting of single  and double traces 
of SU($N$) generators\footnote{The basis 
specified here is that of ref.~\cite{Edison:2012fn}, 
which differs from refs.~\cite{Naculich:2011ep,Naculich:2011my}
by a factor of two in the double-trace terms.}
\ba
T_1 &=&  \Tr(1234) + \Tr(1432),
\qquad\qquad
T_4 =  2 \Tr(13) \Tr(24) , \nn\\
T_2 &=& \Tr(1243) + \Tr(1342),
\qquad\qquad
T_5=  2 \Tr(14) \Tr(23)  ,
\qquad\qquad 
\label{fourpointbasis}  \\
T_3 &=&  \Tr(1324) + \Tr(1423),
\qquad\qquad
T_6 =  2 \Tr(12) \Tr(34) . \nn
\ea
The six-dimensional ket $\ket{A}$ then consists of the coefficients $A_\lam$
of $T_\lambda$ in the amplitude (\ref{basis}).
In this basis, 
the (subleading-color piece of the) one-loop 
soft anomalous dimension matrix takes the form 
\be
\bGam^\One_{\rm sub}
= {2 \over N} 
\left( \begin{array}{cc} 0 & b \\
c & 0
\end{array}
\right),
\qquad\qquad
b = 
\left( 
\begin{array}{ccc}
0 & -2Y & 2X \\ 
2Z &  0 &- 2X \\ 
-2Z&  2Y & 0  
\end{array}
\right) ,
\qquad
c =
\left( 
\begin{array}{ccc}
0 & -X & Y \\
 X & 0 & - Z \\
-Y & Z & 0
\end{array}
\right)
\qquad
\ee
where
\be
X =\log \left(t \over u\right), \qquad
Y =\log \left(u \over s\right), \qquad
Z =\log \left(s \over t\right).
\label{XYZ}
\ee
Hence, \eqn{YMoneloopexact} becomes
\be
\ket{A} \bigg|_{\rm most-subleading-color}
=  \exp \left[ 
{\as \over N \epsilon} 
\left( \begin{array}{cc} 0 & b \\ c & 0 \end{array} \right)
\right]
\ket{H(\eps)} \bigg|_{\rm most-subleading-color} \,.
\ee
Expanding both sides in a loop expansion, we can write 
\be
\ket{A^\EllL } 
= \sum_{\ell=0}^L   
{1 \over (L-\ell)!\epsilon^{L-\ell}}
\left( \begin{array}{cc} 0 & b \\ c & 0 \end{array} \right)^{L-\ell}
\ket{H^{(\ell,\ell)}(\eps)},
\label{msc}
\ee
analogous to \eqn{Lgrav} for gravitational amplitudes.
This expression, valid to all orders in the $\epsilon$ expansion,
was previously obtained in ref.~\cite{Naculich:2009cv} 
for $\cN=4$ SYM theory, 
but here we see that it remains valid for the
four-gluon amplitude in a general gauge theory
provided the dipole conjecture holds.

We rewrite \eqn{msc} separately for even- and odd-loop cases:
\ba
\ket{A^\Tete}
&=& 
\sum_{k=0}^{\ell-1}
{ b (cb)^{\ell-k-1}  \over 
(2\ell-2k)! ~ \epsilon^{2\ell-2k} }
\Biggl[  c
\ket{H^\Tktk(\eps)}
+ 
{ (2\ell-2k) \epsilon}
\ket{H^\Tkptkp(\eps)}
\Biggr]
+
\ket{H^\Tete}, 
\qquad
\label{msceven}
\\
\ket{A^\Teptep } 
&=&
\sum_{ k = 0 }^{\ell}
{ (cb)^{\ell-k}  \over
(2\ell-2k +1)! \epsilon^{2\ell-2k +1} }
\biggl[   c
\ket{H^\Tktk(\eps)}
+ 
(2\ell-2k+1) \epsilon 
\ket{H^\Tkptkp(\eps)}
\biggr],
\label{mscodd}
\ea
where now the kets denote 3-dimensional vectors
\be
\ket{A^\Tete} = 
\left(
\begin{array}{c}
A_1^\Tete  \\
A_2^\Tete  \\
A_3^\Tete  \\
\end{array}
\right),
\qquad
\ket{A^\Teptep} = 
\left(
\begin{array}{c}
A_4^\Teptep  \\
A_5^\Teptep  \\
A_6^\Teptep  \\
\end{array}
\right) .
\ee
We are able to further simplify the expressions (\ref{msceven}) and (\ref{mscodd})
by using the group-theory constraints satisfied by the four-point
amplitude \cite{Naculich:2011ep,Edison:2012fn}.
For $L= 2 \ell$ even,
there are four independent group-theory relations\footnote{These 
relations differ 
from those given in ref.~\cite{Naculich:2011ep}
by some factors of two, due to the change in the trace basis.
\label{fn}}
(only one when $L=0$)
\ba
A_1^\Tete &=&
- \fourthirds A_4^\Tetemo  + \twothirds A_5^\Tetemo + \twothirds A_6^\Tetemo  \,,
\nn\\
A_2^\Tete &=&
+\twothirds A_4^\Tetemo  - \fourthirds A_5^\Tetemo +\twothirds A_6^\Tetemo  \,,
\\
A_3^\Tete &=&
+\twothirds A_4^\Tetemo + \twothirds A_5^\Tetemo -\fourthirds A_6^\Tetemo  \,,
\nn\\
0&=& A_1^\Tetemt + A_2^\Tetemt + A_3^\Tetemt 
- \third \left[A_4^\Tetemo + A_5^\Tetemo + A_6^\Tetemo \right]
\nn\ea
which implies
\be
A_1^\Tete +A_2^\Tete +A_3^\Tete =0
\label{sumvanishes}
\ee
and similarly for the IR-finite $H_\lambda^\Tete$.

For $L=2\ell+1$ odd, we also have four independent 
relations${}^{\ref{fn}}$ (only three when $L=1$)
\ba
A_4^\Teptep &=& A_1^\Tepte +A_2^\Tepte +A_3^\Tepte  \,,
\nn\\
A_5^\Teptep &=& A_1^\Tepte +A_2^\Tepte +A_3^\Tepte  \,,
\\
A_6^\Teptep &=& A_1^\Tepte +A_2^\Tepte +A_3^\Tepte \,,
\nn\\
0 &=& 
A_1^\Teptemt + A_2^\Teptemt + A_3^\Teptemt - \third 
\left[A_4^\Teptemo + A_5^\Teptemo + A_6^\Teptemo \right]
\nn\\
&& + \third 
\left[A_1^\Tepte + A_2^\Tepte + A_3^\Tepte \right]
\nn
\ea
which implies
\be
A_4^\Teptep = A_5^\Teptep = A_6^\Teptep 
\label{equality}
\ee
and similarly for the IR-finite $H_\lambda^\Teptep$.

By virtue of \eqns{sumvanishes}{equality}, together with
\eqn{XYZ}, one can show that
the entries of both 
$c \ket{H^\Tktk}$ and $ \ket{H^\Tkptkp}$ are all equal
\be
c \ket{H^\Tktk} = (Y H_3^\Tktk - X H_2^\Tktk)  \ones , 
\qquad\qquad
\ket{H^\Tkptkp} = H_4^\Tkptkp \ones .
\ee
These together with 
\be
cb \ones = 
(2X^2 + 2Y^2 + 2Z^2) \ones,
\qquad \qquad
b \ones = 
2 \left(\begin{array}{c} X-Y \\ Z-X\\ Y-Z\\ \end{array}
\right)
\ee
allow one to write the even- and odd-loop most-subleading-color amplitude 
(\ref{msceven}) and (\ref{mscodd}) as
\ba
\left(
\begin{array}{c}
A_1^\Tete  \\
A_2^\Tete  \\
A_3^\Tete  \\
\end{array}
\right)
&=&
  \sum_{ k = 0 }^{\ell-1}
{ 2 (2X^2 + 2Y^2 + 2Z^2)^{\ell-k-1} \over
 (2\ell-2k)! \epsilon^{2\ell-2k} }
\nn\\
&\times& \left[ (Y H_3^\Tktk - X H_2^\Tktk)  
+ (2\ell-2k) \epsilon  H_4^\Tkptkp 
\right] 
\left(\begin{array}{c} X-Y \\ Z-X\\ Y-Z\\ \end{array} \right)
+
\left(
\begin{array}{c}
H_1^\Tete  \\
H_2^\Tete  \\
H_3^\Tete  \\
\end{array}
\right) \,,
\nn\\
\label{mscevenfinal}
\ea
\ba
\left(
\begin{array}{c}
A_4^\Teptep  \\
A_5^\Teptep  \\
A_6^\Teptep  \\
\end{array}
\right)
&=&
\sum_{ k = 0 }^{\ell}
{(2X^2 + 2Y^2 + 2Z^2)^{\ell-k} \over
(2\ell-2k +1)!  \epsilon^{2\ell-2k +1}}
\nn\\
&\times & \bigg[ 
(Y H_3^\Tktk - X H_2^\Tktk)  
+ (2\ell-2k+1) \epsilon H_4^\Tkptkp 
\bigg] 
\ones\,.
\qquad\qquad\qquad
\label{mscoddfinal}
\ea
Provided the dipole conjecture holds, 
these expressions give the complete IR-divergent contribution 
to the most-subleading-color $L$-loop four-point amplitudes 
in terms of the IR-finite parts of lower-loop amplitudes, 
as in the case of gravitational amplitudes (\ref{Lgrav}).

Finally we turn our attention to the two leading IR divergences 
of the most-subleading-color amplitudes.
These are given by the $k=0$ terms in \eqns{mscevenfinal}{mscoddfinal}, 
\be
\left(
\begin{array}{c}
A_1^\Tete  \\
A_2^\Tete  \\
A_3^\Tete  \\
\end{array}
\right)
=
  { 2 (2X^2 + 2Y^2 + 2Z^2)^{\ell-1} \over
 (2\ell)! \epsilon^{2\ell} }
\left[ (Y A_3^\Zero - X A_2^\Zero)  
+ (2\ell) \epsilon  H_4^\Oneone
\right] 
\left(\begin{array}{c} X-Y \\ Z-X\\ Y-Z\\ \end{array} \right)
+ \cO(1/\eps^{2 \ell-2}) \,,
\ee
\be
\left(
\begin{array}{c}
A_4^\Teptep  \\
A_5^\Teptep  \\
A_6^\Teptep  \\
\end{array}
\right)
=
{(2X^2 + 2Y^2 + 2Z^2)^{\ell} \over
(2\ell+1)!  \epsilon^{2\ell +1}}
\bigg[ 
(Y A_3^\Zero - X A_2^\Zero)
+ {(2\ell+1) \epsilon} H_4^\Oneone
\bigg] 
\ones
+ \cO(1/\eps^{2 \ell-1}) \,.
\ee
Using the fact that 
\ba
\left(
\begin{array}{c}
A_4^\Oneone (\epsilon) \\
A_5^\Oneone (\epsilon) \\
A_6^\Oneone (\epsilon) \\
\end{array}
\right)
&=&
\left[ {Y A_3^\Zero - X A_2^\Zero \over \epsilon}
+ H_4^\Oneone
\right] 
\ones
\label{mscone}
\ea
we see that the two leading IR divergences can be 
expressed in terms of the one-loop subleading-color amplitude 
evaluated in $D=4-2L\eps$ dimensions:
\ba
\left(
\begin{array}{c}
A_1^\Tete (\epsilon) \\
A_2^\Tete (\epsilon) \\
A_3^\Tete (\epsilon) \\
\end{array}
\right)
&=&
  { 2  (2X^2 + 2Y^2 + 2Z^2)^{\ell-1} \over
 (2\ell-1)! \epsilon^{2\ell-1} }
A_4^\Oneone ((2\ell)\epsilon) 
\left(\begin{array}{c} X-Y \\ Z-X\\ Y-Z\\ \end{array} \right)
+ \cO(1/\eps^{2 \ell-2}) \,,
\qquad\label{evenLYM}
\\
\left(
\begin{array}{c}
A_4^\Teptep (\epsilon)  \\
A_5^\Teptep (\epsilon) \\
A_6^\Teptep (\epsilon) \\
\end{array}
\right)
&=&
{(2X^2 + 2Y^2 + 2Z^2)^{\ell} \over
(2\ell)!  \epsilon^{2\ell}}
A_4^\Oneone ((2\ell+1)\epsilon) 
\ones 
+ \cO(1/\eps^{2 \ell-1}) \,.\label{oddLYM}
\ea
In the case of $\cN=4$ SYM theory,
these relations were previously conjectured in ref.~\cite{Naculich:2008ys}
and proved in ref.~\cite{Naculich:2009cv}.
Here we point out that this is another point of
similarity with gravity amplitudes, which obey the analogous
\eqn{Lloopproponeloop}.

\section{Possible three-loop corrections to the dipole conjecture}
\label{sect-correction}

The results of the previous two sections were contingent
on the validity of the dipole formula for the 
soft anomalous dimension matrix.
The dipole formula holds through at least two loops \cite{Aybat:2006mz},
but could break down beginning at three loops.
Possible forms of a three-loop correction term were considered
in refs.~\cite{Becher:2009qa,Dixon:2009ur},
including a term of the form
\be
\Delta \bGam^\Three =
{1 \over N^3} \bT^a_1 \bT^b_2 \bT^c_3 \bT^d_4 
\left[ f^{ade} f^{cbe}  P_t (s_{ij}) 
+  f^{cae} f^{dbe} P_u (s_{ij})
+ f^{bae} f^{cde}  P_s (s_{ij})\right]
\label{corr}
\ee
which might be generated by the purely gluonic diagram shown in 
fig. 1 of ref.~\cite{Dixon:2009ur}, 
and which contributes at $\cO(1/\eps)$ \cite{Becher:2009qa}.
Other correction terms, 
involving $d^{abc}$, were also discussed in ref.~\cite{Dixon:2009ur}.
In refs.~\cite{Becher:2009qa,Dixon:2009ur,DelDuca:2011ae,Ahrens:2012qz},
strong constraints were put on the 
possible kinematical dependence of the functions $P (s_{ij})$
appearing in \eqn{corr}.
In this section, we discuss the effect of a term of the form (\ref{corr})
on the most-subleading-color four-point amplitude $A^\Threethree$.

Acting with $\Delta \bGam^\Three $ on the 
four-point trace basis (\ref{fourpointbasis}),
as in \eqn{action}, we extract the matrix
\be
\Delta \bGam^\Three =
{1 \over N^3} \left( \begin{array}{cc} a & b \\
c & d
\end{array}
\right)
\ee
where 
\ba
a &=& 
\left(
\begin{array}{ccc}
 0 & 2 N (3 {P_t}-{P_s}-2 {P_u}) & 2 N (2 {P_u}- 3 {P_s}+{P_t}) \\
 2 N ({P_s}+2 {P_t}-3 {P_u}) & 0 & 2 N (3 {P_s}-2 {P_t}-{P_u}) \\
 2 N (3 {P_u}-2 {P_s}-{P_t}) & 2 N (2 {P_s}-3 {P_t}+{P_u}) & 0 \\
\end{array}
\right)\,,
\nn\\
b &=&
\left(
\begin{array}{ccc}
 8 ({P_t}-{P_s}) &
  2 N^2 ({P_t}-{P_s} ) + 4( P_u- {P_s}) 
& 2 N^2 ( P_t - P_s ) + 4 (P_t - P_u) \\
 2 N^2 ({P_s} -{P_u} )  + 4 ({P_s}- {P_t}) 
& 8 ({P_s}-{P_u}) 
& 2 N^2 ({P_s} -{P_u}) + 4 (  {P_t}- {P_u}) \\
 2N^2  (P_u-{P_t} ) + 4 ( {P_s}- {P_t}) 
& 2N^2 ({P_u}-{P_t} ) +4 (P_u - {P_s}) 
& 8 ({P_u}-{P_t}) \\
\end{array}
\right)\,,
\nn
\ea
\ba
c &=&
\left(
\begin{array}{ccc}
 2 ({P_t}-{P_s}) & (N^2+2) ({P_s}-{P_u}) & (N^2+2) ({P_u}-{P_t}) \\
 (N^2+2) ({P_t}-{P_s}) & 2 ({P_s}-{P_u}) & (N^2+2) ({P_u}-{P_t}) \\
 (N^2+2) ({P_t}-{P_s}) & (N^2+2) ({P_s}-{P_u}) & 2 ({P_u}-{P_t}) \\
\end{array}
\right)\,,
\nn\\
d &=&
\left(
\begin{array}{ccc}
 6 N ({P_s}-{P_t}) & 0 & 0 \\
 0 & 6 N ({P_u}-{P_s}) & 0 \\
 0 & 0 & 6 N ({P_t}-{P_u}) \\
\end{array}
\right)\,.
\ea
One can see that $\Delta \bGam^\Three$ is subleading in 
the $1/N$ expansion 
and hence cannot contribute to the planar amplitude $A^\Threezero$.
However, $\cO(1/\eps)$ 
corrections to all the subleading-color amplitudes
$A^\Threeone$, $A^\Threetwo$, and $A^\Threethree$
are possible.\footnote{In appendix B of ref.~\cite{Becher:2009cu},
it was stated that the three-loop correction term (\ref{corr})
contributes at $\cO(N)$.
This is indeed true for the matrix element 
$a$ connecting single-trace terms.
However, the off-diagonal matrix elements $b, c$,
which connect single- and double-trace terms,
have an $\cO(N^2)$ contribution,
as the authors of ref.~\cite{Becher:2009cu} have 
confirmed (private communication). }
In particular,
by keeping only the most-subleading-color contribution of 
$(\as^3/\eps) \Delta \bGam^\Three \ket{A^\Zero} $,
where 
\be
\ket{A^\Zero}= - {4iK \over s t u} 
\left(
\begin{array}{c}
u \\ t \\ s
\end{array}
\right) 
\label{treelevel}
\ee
is the tree-level amplitude,
we obtain the following three-loop contribution to the most-subleading-color amplitude
\be
\left(
\begin{array}{c}
\Delta A_4^\Threethree \\
\Delta A_5^\Threethree \\
\Delta A_6^\Threethree \\
\end{array}
\right)
=
-{8i  K\over \eps}  
{ [ (u-s) P_t + (s-t) P_u + (t-u) P_s  ]  \over 
s t u }  \ones 
+ \cO\left( \eps^0 \right) \,.
\ee
This is an example of a possible IR-divergent contribution 
to the most-subleading-color amplitude that does not arise
from the exponentiation of the one-loop divergence.
Hence, if the dipole formula is modified by a term of the form
(\ref{corr}), then the one-loop-exactness of this class of
amplitudes breaks down.

\section{$L$-loop supergravity/SYM relations}
\label{sect-symsugra}

In the previous sections, we saw that gravity 
and most-subleading-color gauge-theory amplitudes are
one-loop exact, i.e. higher-loop divergences can be 
expressed in terms of one-loop divergences.
In this section, we use this result to derive a relation 
between the two leading divergences 
of the $L$-loop four-point $\cN=8$ supergravity amplitude 
and the most-subleading-color $\cN=4$ SYM amplitudes.

An exact relation between 
the one-loop four-point $\cN=8$ supergravity 
and subleading-color $\cN=4$ SYM amplitudes has long been known
\cite{Green:1982sw, Naculich:2008ys,Naculich:2011my}.
In the notation of the current paper, this relation is
\be
M_4^\One (\eps) = \left( - {\lambda\over 8 \pi^2 } \right)  { A^\Oneone(\eps) \over (A_1^\Zero/u) } 
\label{onelooprelation}
\ee
where $A^\Oneone$ refers to any of 
the four-point subleading-color amplitudes
$A_4^\Oneone = A_5^\Oneone = A_6^\Oneone$,
and we recall that the tree-level amplitude $A^\Zero$ 
is given by \eqn{treelevel}.

In \eqn{Llooponeloop}
we showed that the two leading IR divergences of the 
$L$-loop four-point supergravity amplitude 
can be expressed in terms of the one-loop supergravity amplitude.
In \eqns{evenLYM}{oddLYM},
we derived similar expressions for the two leading IR divergences 
of the $L$-loop most-subleading-color four-point SYM amplitudes.
Combining these with \eqn{onelooprelation}, 
we obtain for odd $L=2\ell+1$ the relation
\ba
M_4^{(2\ell+1)} (\eps) 
&=& \left( - {\lambda\over 8 \pi^2 } \right)^{2\ell+1}
 { (sY-tX)^{2\ell} \over (2X^2 + 2Y^2 + 2Z^2)^\ell }  
\quad  { A^\Ellodd  (\eps) \over (A_1^\Zero/u) } 
+\cO \left( \frac{1}{\epsilon^{2\ell-1}}\right)
\label{oddlooprelation}
\\
&=& \left( - {\lambda\over 8 \pi^2 } \right)^{2\ell+1}
\left[\frac{(s\log s +t\log t+u\log u)^2}{2(\log^2(t/u)+\log^2(u/s)+\log^2(s/t))}\right]^\ell
{ A^\Ellodd  (\eps) \over (A_1^\Zero/u) } 
+\cO \left( \frac{1}{\epsilon^{2\ell-1}}\right)\,.
\nn
\ea
Again,
$A^\Ellodd$ refers to any of 
the most-subleading-color four-point amplitudes
$A_4^\Ellodd= A_5^\Ellodd=A_6^\Ellodd$
(cf. \eqn{equality}).

For even $L= 2\ell$, a similar relation holds, namely
\be
M_4^{(2\ell)} (\eps) 
= \left(  {\lambda\over 8 \pi^2 } \right)^{2\ell}
 { (sY-tX)^{2\ell-1} \over (2X^2 + 2Y^2 + 2Z^2)^{\ell-1} }  
\quad { A_1^\Elleven(\eps)  \over 2 (X-Y) (A_1^\Zero/u) }
+\cO \left( \frac{1}{\epsilon^{2\ell-2}}\right)
\label{evenlooprelation}
\ee
for $\ell\geq 1$.
The factor $A^\Elleven_{1}/(X-Y)$ can of course be replaced
with $A^\Elleven_{2}/(Z-X)$ or $A^\Elleven_{3}/(Y-Z)$,
or in fact with $\left(  A_1^\Elleven - A_2^\Elleven \right)/3X$
(since $X+Y+Z=0$) giving
\be
M_4^{(2\ell)} (\eps) 
= \left(  {\lambda\over 8 \pi^2 } \right)^{2\ell}
 { (sY-tX)^{2\ell-1} \over (2X^2 + 2Y^2 + 2Z^2)^{\ell-1} } 
{\left(  A_1^\Elleven (\eps) - A_2^\Elleven (\eps) \right)
\over 6X (A_1^\Zero/u) }
+\cO \left( \frac{1}{\epsilon^{2\ell-2}}\right) \,.
\ee
To repeat, these relations are immediate consequences of 
eqs.~(\ref{Llooponeloop}), (\ref{evenLYM}), (\ref{oddLYM}),
and (\ref{onelooprelation}).

An interesting question is whether the relations (\ref{oddlooprelation})
and (\ref{evenlooprelation}) remain valid beyond the leading
two orders in the Laurent expansion.  
Unfortunately, the answer will turn out to be no.

To see this, observe that for $L=2$, \eqn{evenlooprelation}
states that 
\be
M_4^\Two (\eps) 
= \left(  {\lambda\over 8 \pi^2 } \right)^{2}
 {  (sY-tX)  \over 2 (X-Y) }
 {  A_1^\Twotwo (\eps)  \over (A_1^\Zero/u)  }
+\cO \left( \eps^0 \right).
\label{twolooprelation}
\ee
We know this to be valid at $\cO(1/\eps^2)$ and $\cO(1/\eps)$,
and the question is whether it continues to hold at  $\cO(\eps^0)$.
To answer this, we recall the {\it exact} two-loop supergravity/SYM relation derived
in \cite{Naculich:2008ys,Naculich:2011my}
\be
M_4^\Two (\eps) 
= \left(  {\lambda\over 8 \pi^2 } \right)^{2}
{\left( u A_1^\Twotwo (\eps) +t A_2^\Twotwo (\eps) +s A_3^\Twotwo (\eps) \right)
\over 6 (A_1^\Zero/u) }   \,.
\label{exactrelation}
\ee
A short calculation using 
$s+t+u=0$, $X+Y+Z=0$,  and \eqn{sumvanishes} 
shows that \eqns{twolooprelation}{exactrelation} are consistent provided that
\be
{A_1^\Twotwo \over X-Y} 
\stackrel{?}{=}
{A_2^\Twotwo \over Z-X} 
\overset{?}{=}
{A_3^\Twotwo \over Y-Z}  \,.
\label{condition}
\ee
In fact, were \eqn{condition} to hold, then \eqn{oddlooprelation} 
would also hold at $\cO(1/\eps)$ for $L=3$ 
(provided that the dipole conjecture is also valid at three loops),
and in fact for the $\cO(1/\eps^{L-2})$ term at higher loops as well.

While \eqn{condition} evidently holds for the IR-divergent parts of 
the amplitude (cf. \eqn{mscevenfinal}), we have verified that it
fails at $\cO(\eps^0)$, that is
\be
(Z-X) A^\Twotwo_1 -  (X-Y) A^\Twotwo_2 \neq 0
\label{notequal}
\ee
using the explicit expressions for 
the two-loop most-subleading-color $\cN=4$ SYM four-point
amplitudes (see appendix A). 
To ensure that the complicated expression 
obtained for the left hand side of \eqn{notequal}
does not vanish due to polylogarithmic identities,
we evaluated it numerically for various values of the kinematic 
variables, obtaining nonzero results.
Finally, we checked that the symbol \cite{Goncharov:2010jf} 
for the expression on the left hand side of \eqn{notequal} 
does not vanish (sometimes a non-obvious polylog identity reduces a long expression to a simple one, which can be made explicit
by the calculation of the symbol \cite{DelDuca:2009au,DelDuca:2010zg,Goncharov:2010jf}, and moreover an identity could be valid only up
to terms with zero symbol); see appendices B and C for details.

Consequently, \eqns{oddlooprelation}{evenlooprelation} are valid
for the two leading terms, but break down at the next order in the
Laurent expansion.


\section{Conclusions}
\label{sect-conclusions}

In this paper, we explored parallels between the IR behavior 
of gravitational amplitudes and that of the 
most-subleading-color gauge-theory amplitudes.
Both sets of amplitudes have a leading IR divergence of $\cO(1/\eps^L)$ at $L$ loops, 
due to the absence of collinear divergences. 
We have shown that, if the dipole conjecture for the IR behavior of gauge-theory 
amplitudes is valid, then the most-subleading-color amplitudes,
like gravity amplitudes, are one-loop-exact;  that is,
higher-loop divergences are determined by the one-loop result.
Specifically, the all-loop amplitude is given by the 
exponential of the one-loop soft anomalous dimension matrix $\bGam^\One$
acting on the IR-finite hard function.

Assuming the validity of the dipole conjecture, we computed an expression for 
the complete IR behavior of the $L$-loop most-subleading-color four-point amplitude 
in terms of the finite parts of lower-loop amplitudes.
Similar expressions could be derived for five- and higher-point amplitudes using
the explicit form for the  one-loop soft anomalous dimension matrix $\bGam^\One$.
We note that $\bGam^\One$ is essentially equivalent to the (transpose of the) iterative matrix $G$
specified in refs.~\cite{Naculich:2011ep,Edison:2011ta,Edison:2012fn},
defined by attaching a rung between two external legs $i$ and $j$ of an element of the trace basis $T_\lam$.
In the present context, each rung corresponds to the exchange of a soft gluon between the
corresponding external particles, accompanied by a factor of $\log(\mu^2/-s_{ij})$. 

Corrections to the dipole conjecture may occur at three loops
and beyond, although the possible form of such corrections is highly constrained.
In recent work, Oxburgh and White \cite{Oxburgh:2012zr} use BCJ duality and the
double-copy property to study the IR behavior of gauge theory and gravity.  
They emphasize that the known IR structure of gravity is insensitive
to possible corrections to the dipole conjecture in gauge theories.
Therefore the presence or absence of such corrections at three loops
will likely require a Laurent expansion of the three-loop non-planar diagrams
contributing to the gauge amplitude.
We showed that, though collinear IR divergences remain absent,
corrections could spoil the one-loop-exactness of most-subleading-color amplitudes. 

Finally, we showed that the similarities 
between gravity and most-subleading-color amplitudes
allow us to deduce a relation between
$L$-loop four-point $\cN=8$ supergravity and 
most-subleading-color $\cN=4$ SYM amplitudes 
that holds for the two leading IR divergences,
$\cO(1/\epsilon^L)$ and $\cO(1/\epsilon^{L-1})$, 
but breaks down at $\cO(1/\epsilon^{L-2})$.

\section*{Acknowledgments}

We would like to thank T. Becher and M. Neubert for useful correspondence.
The research of S.~Naculich is supported in part by the
NSF under grant no.~PHY10-67961.
The research of H.~Nastase is supported in part by
CNPQ grant 301219/2010-9.
The research of H.~Schnitzer is supported in part by
the DOE under grant DE-FG02-92ER40706.

\vfil\break
\begin{appendix}

\section{Two-loop most-subleading-color $\cN=4$ SYM four-point amplitude}

The two-loop most-subleading-color four-point amplitudes in $\cN=4$ SYM theory
are given in terms of two-loop planar and nonplanar scalar integrals \cite{Bern:1997nh}.
Explicit expressions for these may be derived employing
the Laurent expansions of the planar \cite{Bern:2005iz}
and non-planar \cite{Tausk:1999vh} integrals.
Analytically continuing these integrals to the kinematic region
$t>0$ and $s$, $u<0$,
we obtain the expression
\ba
{A_1^\Twotwo (\eps) \over (A_1^\Zero/u)}
&=& \left(\mu^2 \over t\right)^{2\eps} \bigg\{
{ \left( - s \log y-u \log (1-y)-i \pi (s+u) \right) (X-Y)
\over
\eps^2}
\\[3mm]
&+&
{\left(2 (s+u)  \log y \log (1-y)  +2 i \pi  u \log y + 2 i \pi  s \log (1-y)\right)
(X-Y)
\over
\eps}
\nn\\ [3mm]
&+&
  \left( -20 s - 4 u \right) S_{3,1} (y)
+ \left( 4 s - 4 u \right) S_{2,2} (y)
+ \left( -8 s - 4 u \right) S_{1,3} (y)
\nn\\&+&
 \big[ \left( 10 s  -4 u\right)  \log y 
+ \left( 8 s +10 u \right) \log (1-y) 
+ \left( 14 i \pi  s + 10 i \pi  u \right) \big] S_{2,1} (y)
\nn\\&+&
\big[  \left( 4 s + 8 u \right) \log y
+ \left( -16 s - 8 u \right) \log (1-y) 
+ \left( -4 i \pi s+  4 i \pi u \right) \big] S_{1,2} (y)
\nn\\&+&
  \big[ 6u \log^2 y 
+ (-8 s - 10 u) \log y \log(1-y)
+ (  -12 s - 6 u) i \pi \log y 
\nn\\&+&
 (8 s - 2 u) i \pi \log (1-y)
+(-4 s + 4 u)\pi^2
\big] S_{1,1}(y)
+\frac{1}{2} s \log ^4y
-\frac{4}{3} (s-u) \log ^3y \log (1-y)
\nn\\&-&
(2s+4u) \log^2y \log ^2(1-y)
+ (4s+4u) \log y \log ^3(1-y)
-u \log ^4(1-y)
\nn\\&-&
i \pi  (s+2u) \log ^3y
+5 i \pi  u \log ^2y \log (1-y)
-4 i \pi  u \log y \log ^2(1-y)
\nn\\&+&
4 i \pi  s \log ^3(1-y)
+ \left( \frac{13}{6}  s -2  u\right)\pi^2 \log^2y
+ \left(  -\frac{13}{3}  s+\frac{37}{6}  u\right) \pi^2\log y \log (1-y)
\nn\\&+&
 \left(-2  s  -\frac{13}{3}  u \right)\pi^2 \log^2(1-y)
+ \left( -2 i \pi ^3 s +\frac{1}{6} i \pi ^3 u -2 s \zeta_3 \right) \log y
\nn\\&+&
\left( -\frac{5}{3} i \pi ^3 s -\frac{19}{6} i \pi ^3 u +8 s \zeta_3 +2 u \zeta_3 \right) \log (1-y)
\nn\\&+&
\frac{17 \pi ^4 s}{15} +\frac{\pi ^4 u}{6} +2 i \pi  s \zeta_3 - 6 i \pi u  \zeta_3 
+ \cO(\eps) \bigg\}
\nn
\ea
where 
$ y \equiv -s/t$ and $S_{n,p}(y)$ denote the generalized polylogarithms of Nielsen \cite{Kolbig}.
In this region, the variables $X$, $Y$, and $Z$ defined in  \eqn{XYZ} become
\ba
X &=& - \log (1-y) -i \pi \,,\nn\\
Y &=& - \log y + \log (1-y) \,,\\
Z &=&   \log y + i \pi \,. \nn
\ea
(See appendix A of ref.~\cite{Naculich:2008ew} for details on the performance of the
analytic continuation.)

We also have
\ba
{A_2^\Twotwo (\eps) \over (A_1^\Zero/u)}
&=& \left(\mu^2 \over t\right)^{2\eps} \bigg\{
{\left(-  s\log y-u \log (1-y)-i \pi (s+u) \right) (Z-X)
\over \eps^2}
\\ [3mm]
&+&
{\left(
2 (s+u) \log y \log (1-y)+2 i \pi  u \log y + 2 i \pi  s \log (1-y)\right) (Z-X)
\over \eps}
\nn\\ [3mm]
&+&
 \left( 16 s - 4 u \right) S_{3,1} (y)
+ \left( -8 s + 8 u \right) S_{2,2} (y)
+ \left( 4 s - 16 u \right) S_{1,3} (y)
\nn\\&+&
\big[ \left( -14 s - 4 u \right) \log y
+ \left( -4 s - 2 u \right) \log (1-y) 
+ \left( -10 i \pi s-14 i  \pi u \right)  \big] S_{2,1} (y)
\nn\\&+&
 \big[ \left( 10 s - 4 u \right) \log y
+ \left( 8 s - 2 u \right) \log (1-y) 
+ \left( 14 i\pi s+10 i\pi u \right) \big] S_{1,2} (y)
\nn\\&+&
  \big[
(6s+6u) \log^2 y + (4s+2u) \log y \log(1-y) + (6s+18u) i \pi \log y 
\nn\\&+&
(-4s+4u) i \pi \log (1-y) +(8s-8u)\pi^2 \big] S_{1,1}(y)
+ \frac{1}{2} s \log ^4(y)
+\frac{2}{3} (s+2u) \log ^3(y) \log (1-y)
\nn\\&-&
(2 s+u)  \log ^2(y) \log ^2(1-y)
-(2 s+u) \log (y) \log ^3(1-y)
+\frac{1}{2} u \log ^4(1-y)
\nn\\&+&
i \pi  (s-2u) \log ^3(y)
+ 5 i \pi  u \log ^2(y) \log (1-y)
-i \pi (6 s+u)  \log (y) \log ^2(1-y)
\nn\\&+&
i \pi  (-2s+u) \log ^3(1-y)
+ \left( \frac{13}{6}  s +4 u \right) \pi ^2 \log ^2 y
+\left( \frac{13}{6}  s -\frac{35}{6}  u\right) \pi^2 \log y \log (1-y)
\nn\\&+&
\left( 4 s +\frac{13}{6} u \right) \pi ^2  \log ^2(1-y)
+\left( \frac{11}{2} i \pi ^3 s +\frac{1}{6} i \pi ^3 u -2 s \zeta_3 \right)  \log y
\nn\\&+&
\left( \frac{5}{6} i \pi ^3 s +\frac{29}{6} i \pi ^3 u -4 s \zeta_3 +2 u \zeta_3\right) \log (1-y)
\nn\\&-&
\frac{28 \pi ^4 s}{15} -\frac{\pi ^4 u}{3} -10 i \pi  s \zeta_3 -6 i \pi  u \zeta_3
+\cO(\eps)
\bigg\} \,.
\nn
\ea
Finally, $A_3^\Twotwo$ is obtained using
\be
A_1^\Twotwo +A_2^\Twotwo + A_3^\Twotwo = 0\,.
\ee
To our knowledge, explicit expressions for these
amplitudes have not appeared previously in the published literature.
The uniform transcendentality of these expressions,
previously noted in ref.~\cite{Naculich:2008ys},
is evident in these expressions,
underlining the fact that this property of $\cN=4$ SYM
observables extends beyond the planar approximation.

Using the expressions above, one may verify that
\be 
(Z-X) A^\Twotwo_1 -  (X-Y) A^\Twotwo_2 \neq 0\,.
\ee
For this purpose, it is simplest to examine the coefficient of $\zeta_3$.

\section{Symbology}

In this appendix we review the general features of symbols. For more details, see refs.~\cite{Duhr:2011zq} and \cite{Gaiotto:2011dt}.

The symbols are simply defined for Goncharov polynomials of one variable, defined recursively as
\be
G(a_1,...,a_n;x)=\int_0^x\frac{dt}{t-a_1}G(a_2,...,a_n;t)
\ee
with 
\be
G(x)=G(;x)=1;G(0)=0
\ee
Other functions are obtained from them as 
\bea
G(\vec{0}_n;x)=\frac{1}{n!}\log^n x\cr
G(\vec{a}_n;x)=\frac{1}{n!}\log^n\left(1-\frac{x}{a}\right)\cr
G(\vec{0}_{n-1},a;x)=-{\rm Li}_n\left(\frac{x}{a}\right)\cr
G(\vec{0}_n,\vec{a}_p;x)=(-1)^pS_{n,p}\left(\frac{x}{a}\right)
\eea
The Goncharov polylogarithms of one variable are similarly defined with the harmonic polylogarithms \cite{Remiddi:1999ew}
$H$ with indices $0$ and $\pm 1$, which are related to the Nielsen polylogarithms by
\be
S_{n,p}(x)=H(\vec{0}_n,\vec{1}_p;x).
\ee
The symbol of a Goncharov polynomial is defined as sum of terms of the type tensor product of $R_i$'s, understood as $d\log R_i=dR_i/R_i$'s, i.e. such that 
the rules the $R_i$'s satisfy follow from this $d\log$ form. These tensor monomials are written as $R_1\otimes ...\otimes R_n$ and satisfy
\bea
&& ...\otimes (R_1\cdot R_2)\otimes ...=...\otimes R_1\otimes ...+...\otimes R_2 \otimes ...\cr
&& \Rightarrow \otimes (R_1)^n\otimes =n...\otimes R_1\otimes ...\cr
&& ...\otimes cR_1\otimes ...=...\otimes R_1\otimes ...\cr
&&...\otimes c\otimes...=0\cr
&&\Rightarrow ...\otimes R_1\otimes ...=-...\otimes 1/R_1\otimes...
\eea
where $R_1,R_2,...$ are variable monomials and $c$ is a constant. The symbol of an object $T_k$, a priori a function of several variables, an 
extension of the simple Goncharov polynomials of one variable above and 
defined recursively as 
\be
T_k=\int_a^b d\log R_1\circ ...\circ d\log R_n=\int_a^b\left(\int_a^t d\log R_1\circ...d\log R_{n-1}\right)d\log R_n(t), 
\ee
is 
\be
S[T_k]=R_1\otimes R_2\otimes ...\otimes R_n.
\ee
From this definition we obtain immediately the symbol of a $Li_k$ polylogarithm,
\be
S[{\rm Li}_k(z)]=-(1-z)\otimes z\otimes...z
\ee
(there are $k-1$ factors of $z$), as a particular case of the Goncharov polylogarithms. 
Note that $(1-z)$ and $(z-1)$ are the same, since they differ by multiplication by the constant $-1$, however, the 
overall minus sign is for the tensor monomial, it does not belong into any of the tensored factors. 

We can also define the rule for multiplication of two symbol terms $S[F]=\otimes_{i=1}^n R_i$ and $S[G]=\otimes _{i=n+1}^m R_i$, as
\be
S[FG]=\sum_\Pi \otimes _{i=1}^{m+n}R_{\Pi(i)}
\ee
where the permutations $\Pi$ preserve the original order of the factors in $S[F]$ and in $S[G]$ within $S[FG]$. For example, if $n=m=2$ we get
\bea
S[FG]&=&R_1\otimes R_2\otimes R_3\otimes R_4+R_1\otimes R_3\otimes R_2\otimes R_4+R_1\otimes R_3\otimes R_4\otimes R_2\cr
&&+R_3\otimes R_1\otimes R_2\otimes R_4+R_3\otimes R_1\otimes R_4\otimes R_2+R_3\otimes R_4\otimes R_1\otimes R_2.
\eea

For logs and their products we obtain
\bea
&&S[\log x]=x\cr
&&S[\log x\log y]=x\otimes y+y\otimes x.
\eea

Finally, for the Nielsen polylogarithms
\be
S_{n,p}=\frac{(-1)^{n+p-1}}{(n-1)!p!}\int_0^1dt \frac{\log^{n-1}(t)\log^p(1-xt)}{t},
\ee
the symbol is given by
\be
S[S_{n,p}(x)=H(\vec{0}_n,\vec{1}_p;x)]=(-1)^p(1-x)\otimes(1-x)\otimes ...\otimes (1-x)\otimes x\otimes x...\otimes x
\ee
where there are $p$ $1-x$'s and $n$ $x$'s. 

\section{Symbol relation}

In this appendix, we describe how we tested the relation (\ref{notequal})
using the symbol. 
Appendix B reviews some salient features of symbols of polylogarithms.

The amplitude $A_1^{(2,2)}$ from Appendix A, divided by $A_1^\Zero/u$, has 
terms proportional to the independent variables $s$ and $u$ (where $s+t+u=0$), 
and these kinematic factors are not touched by the symbol. Therefore we will only check the $s$-terms in the desired relation,
\be
(Z-X) A_{1}^{(2,2)}
\stackrel{?}{=}
(X-Y)A_{2}^{(2,2)},\label{a22identity}
\ee
for the finite order pieces.
(We know the IR divergent pieces satisfy this relation, 
and we have in fact explicitly checked this.)

Given that we are interested only in the relation between symbols, the analytical continuations become simpler. To find the first cyclic term, in 
the $t>0, s,u<0$ region, we need to first analytically continue to $s>0, t,u<0$ and then do the cyclic shift. The analytical continuation gives
\bea
&& y=-\frac{s}{t}\rightarrow y e^{-2\pi i}\Rightarrow \log y \rightarrow \log y -2\pi i\cr
&& 1-y=-\frac{u}{t}\rightarrow -(1-y)e^{-\pi i}\cr
&&{\rm Li}_k(y)\rightarrow {\rm Li}_k(y e^{-2\pi i})={\rm Li}_k(y)\cr
&&S_{1,k}(y)\rightarrow S_{1,k}(ye^{-2\pi i})=S_{1,k}(y)
\eea
and then the change $(s,t,u)$ into $(t,u,s)$ leads to $y\rightarrow 1/(1-y)$ and $1-y \rightarrow -y/(1-y)$. All in all, we obtain
\bea
&&T\rightarrow V+2\pi i\cr
&& V\rightarrow -T-V-\pi i \cr
&& {\rm Li}_k(y)\rightarrow {\rm Li}_k\left(\frac{1}{1-y}\right)\cr
&& S_{1,k}(y)\rightarrow S_{1,k}\left(\frac{1}{1-y}\right).
\eea
To find the second cyclic term in the $t>0$, $s,u<0$ region,  we first analytically continue to $u>0, s,t<0$, and then do the cyclic shift. The analytical continuation gives
\bea
&&y\rightarrow -y e^{-i\pi}\Rightarrow \log y \rightarrow \log (-y) -\pi i\cr
&&1-y \rightarrow (1-y)e^{-2\pi i}\cr
&&{\rm Li}_k(y)\rightarrow {\rm Li}_k(y)+{\rm terms\;\; of \;\; 0 \;\; symbol}\cr
&&S_{1,k}(y)\rightarrow S_{1,k}(y)+{\rm terms\;\; of \;\; 0 \;\; symbol}
\eea
and then the change $(s,t,u)$ into $(u,s,t)$ leads to $y\rightarrow -(1-y)/y$ and $1-y\rightarrow 1/y$. All in all, we obtain 
\bea
&&T\rightarrow -V -T +\pi i\cr
&& V\rightarrow T-2\pi i\cr
&&{\rm Li}_k(y)\rightarrow {\rm Li}_k\left(-\frac{1-y}{y}\right)+{\rm terms\;\; of \;\; 0 \;\; symbol}\cr
&& S_{1,k}(y)\rightarrow S_{1,k}\left(-\frac{1-y}{y}\right)+{\rm terms\;\; of \;\; 0 \;\; symbol}
\eea
and now we can ignore the terms with zero symbol, involving transcendental constants like $\pi$. That means that ignoring these terms, the relation we 
need to check is 
\be
A_{1}^{(2,2)}(\log y +\log (1-y))=A_{2}^{(2,2)}(\log y -2 \log (1-y))+{\rm terms\;\; of \;\; 0 \;\; symbol}
\ee
and as we mentioned, we will only check the $s$-terms. 

The resulting symbol contains tensor products of $y$ and $(1-y)$ monomials forming a 5-fold tensor product, so there are $2^5=32$ independent 
tensor structures which should have zero coefficient if this identity is to hold in symbol. We have checked 4 of these coefficients, and shown them to 
be nonzero. 

In conclusion, the identity (\ref{a22identity}), and therefore also the SYM-supergravity relation at two-loops, does not hold to finite order, not even 
in symbol.

\end{appendix}

\end{document}